\documentstyle[12pt]{article}
\newlength{\extraspace}
\newlength{\extraspaces}
\def\numberbysection{\@addtoreset{equation}{section}
\def\theequation{\arabic{section}.\arabic{equation}}}

\newcommand{\be}{\begin{equation}
\addtolength{\abovedisplayskip}{\extraspaces}
\addtolength{\belowdisplayskip}{\extraspaces}
\addtolength{\abovedisplayshortskip}{\extraspace}
\addtolength{\belowdisplayshortskip}{\extraspace}}
\newcommand{\ee}{\end{equation}}
\newcommand{\ba}{\begin{eqnarray}
\addtolength{\abovedisplayskip}{\extraspaces}
\addtolength{\belowdisplayskip}{\extraspaces}
\addtolength{\abovedisplayshortskip}{\extraspace}
\addtolength{\belowdisplayshortskip}{\extraspace}}
\newcommand{\ea}{\end{eqnarray}}
\begin{document}
\addtolength{\baselineskip}{.7mm}
\thispagestyle{empty}
\begin{flushright}
TIT/HEP--267 \\
NUP-A-94-18 \\
hep-th/9410090 \\
October, 1994
\end{flushright}
\vspace{2mm}
\begin{center}
{\large{\bf
QUANTUM DIFFUSION PROCESS WITH A COMPLEX WEIGHT
ON THREE DIMENSIONAL LATTICE}}\footnote{
An updated version of the talk presented by C. Itoi
in The International Seminar Devoted to the 140th Birthday of
Henri Poincar\'e,
IHPE, Protovino, Russia, 27/June - 1/July, 1994, to appear in
the Proceedings.}

%
{\sc Masako Asano}
\\[2mm]
{\it Department of Physics, Tokyo Institute of Technology, \\[1mm]
Oh-okayama, Meguro, Tokyo 152, Japan} \\[3mm]
{\sc Chigak Itoi}
\\[2mm]
{\it Department of Physics, \\[1mm]
College of Science and Technology, Nihon University, \\[1mm]
Kanda Surugadai, Chiyoda, Tokyo 101, Japan} \\[1mm]
and \\[1mm]
{\sc Shin-Ichi Kojima}
\\[2mm]
{\it Department of Physics,
Tokyo Institute of Technology, \\[1mm]
Oh-okayama, Meguro, Tokyo 152, Japan} \\ [4mm]
{\bf Abstract}\\[2mm]

{\parbox{13cm}{\hspace{5mm}
A rigorous definition of a path integral
for a spinning particle in three dimensions is given
on a regular cubic lattice.
The critical diffusion
constant and the associated critical exponents in each spin
are calculated.
Continuum field theories such as Klein-Gordon,
Dirac and massive Chern-Simons theories are
constructed near these critical points.
The universality of obtained results is argued on some
other lattices.}}
\end{center}

%
\setcounter{section}{0}
\setcounter{equation}{0}
%
%
\section{Introduction}
A path integral representation of spinning particles
in three dimensional
Euclidian space
is studied extensively
\cite{P} - \cite{T}.
The free energy of spinning particles with spin $J \in {\bf Z}/2$
can be expressed in terms of a bosonic path integral.
It is well-known that the critical behavior of this model differs
from that in Brownian motion.
Since a path has to
pay energy to bend itself due to the spin factor,
a typical path in this model is much smoother
than that in Brownian motion. And then, the spin factor affects
the critical exponents of the model.
This path integral representation has been used
to argue a spinor field coupled to Chern-Simons field in
\cite{P} \cite{I} \cite{T}.
However, it is difficult to check the self-consistency among
several heuristic limitation procedures
employed to define the path integral, and
therefore the reliability of the
nonperturbative consequence obtained by this method remains unclear.\\

In this talk, we report our recent work on construction of
continuum spinor field theories in terms of random walks
on a three dimensional lattice \cite{CI} \cite{AIK}.
The random walk
is defined on lattice and the continuum limit is discussed
rigorously. This model corresponds to a simple
generalization of the Markovian process for
solving two dimensional Ising model
\cite{V} to that in three dimensions with other spins.
In the case of two dimensional Ising model, the number of
diagrams in the high temperature expansion is
evaluated in terms of a particular Markovian process.
We find a critical point for each value of $J$ and study its critical
behaviors to construct continuum field theories.
In regular cubic lattice,
Klein-Gordon, Dirac and massive Chern-Simons
fields can be constructed at corresponding
critical points. In other lattice such as
triclinic lattice, only Klein-Gordon
and Dirac fields can be constructed.
Therefore, the construction of the
Klein-Gordon and the Dirac theories by this random walk model
is expected to be universal but that of the massive Chern-Simons
theory is not universal. \\


\section{Definition of the model}

 Let us consider the
following
Markovian diffusion process
on a three dimensional lattice
with an amplitude
depending on the preceding step.
To define the free energy,
we evaluate a sum over all paths leaving and returning to
the origin. First, we work on regular cubic lattice.
Let ${\bf x}$ and ${\bf e}_\alpha$ be a
lattice site and a bond indices, respectively.
The length of each bond is taken to be unity.
We define a probability  amplitude
$A^{\alpha \beta}_{t}({\bf x})$
for a particle that starts at the origin, moving
in direction ${\bf e}_\alpha$ to propagate in $t$ steps
to a site ${\bf x}$ moving in direction ${\bf e}_\beta$.
Here, $A^{\alpha \beta}_{t}({\bf x})$
 satisfies a recursion relation
\begin{equation}
A^{\alpha \beta}_{t+1}({\bf x})
= \sum_{\gamma \neq -\beta}
\kappa \ e^{i J S({\bf e}_3, {\bf e}_\beta, {\bf e}_\gamma)}
A^{\alpha \gamma}_t ({\bf x}-{\bf e}_\beta),
\label{rec}
\end{equation}
where $\kappa$ is a diffusion constant and
$J$ is a spin parameter.
The sum is taken over all nearest neighbor bonds ${\bf e}_\gamma$
at the site ${\bf x}$ except ${\bf e}_{-\beta}$,
where we denote ${\bf e}_{-\gamma} =-{\bf e}_{\gamma}$.
In our model any backtracking process is forbidden.
$S({\bf e}_\alpha, {\bf e}_\beta, {\bf e}_\gamma)$
is the area of a spherical triangle with vertices
${\bf e}_\alpha$, ${\bf e}_\beta$ and ${\bf e}_\gamma$ having
an orientation, basically.
Some ambiguous cases are defined by
$S({\bf e}_3, {\bf e}_{-3}, {\bf e}_1) = 0 $,
$S({\bf e}_3, {\bf e}_{-3}, {\bf e}_{-1}) = 2 \pi $ and
$S({\bf e}_3, {\bf e}_{-3}, {\bf e}_{\pm 2}) = \pm \pi$.
The initial condition is given by
\begin{equation}
A^{\alpha \beta}_0 ({\bf x})
 = \delta^{\alpha \beta}
  \delta_{x_1 0}~ \delta_{x_2 0}~ \delta_{x_3 0}.
\label{initial}
\end{equation}
Here, we consider only half integer $J$.
This diffusion process
in two dimensions with $J= \frac{1}{2}$
enables us to calculate all orders in the high temperature
expansion for the Ising model. In this representation,
Majorana spinor is obtained as
a critical theory of the
two dimensional Ising model.
Our three dimensional diffusion model is
regarded as its generalization.
A free energy $f(\kappa, J)$
per unit volume is
\begin{equation}
f(\kappa, J)= - (-1)^{2J} \sum_{t=1} ^\infty
\sum_{\alpha= \pm 1} ^{\pm 3} \frac{1}{t}
A^{\alpha \alpha}_t({\bf 0}).
\label{fe1}
\end{equation}
In momentum space we
obtain the Fourier transformed
recursion relation as
a block diagonal form
\begin{equation}
B^{\alpha \beta}_{t+1}({\bf p})
= \sum_\gamma \kappa B^{\alpha \gamma} _t ({\bf p})
Q^{\gamma \beta} ({\bf p}, J),
\label{frc}
\end{equation}
where $B^{\alpha \beta}_t (\bf p)$ is the Fourier transform
of the amplitude $A^{\alpha \beta} _t (\bf x)$.
The matrix $Q({\bf p}, J)$ has the form
\begin{equation}
 Q({\bf p}, J)=\left[
\matrix{e^{-ip_2} & e^{ip_1 +i \frac{\pi}{2} J}
& 0 & e^{-ip_1 -i \frac{\pi}{2} J}
& e^{-ip_3} & e^{i p_3 -i \pi J} \cr
e^{-ip_2-i \frac{\pi}{2} J} & e^{i p_1}
& e^{ip_2+i \frac{\pi}{2} J} & 0
& e^{-i p_3} & e^{ip_3 -i2 \pi J} \cr
0 & e^{ip_1 -i \frac{\pi}{2} J}
& e^{ip_2} & e^{-ip_1 +i \frac{\pi}{2} J}
& e^{-ip_3} & e^{ip_3+ i \pi J} \cr
e^{-i p_2+i \frac{\pi}{2} J} & 0
& e^{ip_2-i \frac{\pi}{2} J} & e^{-ip_1}
& e^{-ip_3} & e^{ip_3} \cr
e^{-ip_2} & e^{ip_1}
& e^{ip_2} & e^{-ip_1}
& e^{-ip_3} & 0 \cr
e^{-ip_2 + i \pi J} & e^{i p_1 + i 2 \pi J}
& e^{i p_2 - i \pi J} & e^{-i p_1}
& 0 & e^{i p_3} \cr} \right]
\label{mx}
\end{equation}
The solution of the recursion equation (\ref{frc}) with (\ref{initial})
is
\begin{equation}
B_{t}({\bf p}) =
  \{\kappa Q({\bf p}, J) \}^t.
  \end{equation}
 The free energy can be written in terms of
the matrix $Q({\bf p}, J)$
\begin{equation}
f(\kappa, J)= -(-1)^{2J} \int \frac{d^3 p}{(2 \pi)^3}
{\rm Tr} \sum _{t=1} ^{\infty}
\frac{1}{t} \{ \kappa Q({\bf p}, J) \} ^t
\label{sum}
\end{equation}
This summation converges only when $\kappa^{-1}$ is larger than
the maximal absolute value of the eigenvalue of the matrix
$Q({\bf p}, J)$. This can be checked numerically.
In this case, the summation becomes
\begin{equation}
f(\kappa, J)=(-1)^{2J} \int \frac{d^3 p}{(2 \pi)^3}
{\rm Tr} \log[1 - \kappa Q({\bf p}, J)]
\label{fe2}
\end{equation}


\section{Critical Theories}

Physical quantities become non-analytic
at the critical point as a function of the
diffusion constant $\kappa$, due to the finite convergence radius
of the summation (\ref{sum}).
If the correlation length diverges
at the critical point,
we can take
a continuum field theory limit. The model
at several critical points has a
certain continuum limit.
One can check the maximal absolute value in the eigenvalues
of the matrix $Q({\bf p}, J)$ is given at ${\bf p}={\bf 0}$
at least numerically.
 A critical diffusion constant $\kappa_c$
is determined by the eigenvalue equation of
$Q({\bf 0}, J)$
\begin{equation}
\det [\kappa_c ^{-1}- Q({\bf 0}, J)] = 0.
\label{ee}
\end{equation}
At $\kappa=\kappa_c$, the free energy becomes
non-analytic. Since $S({\bf e}_\alpha, {\bf e}_\beta,
{\bf e}_\gamma)$ in (\ref{rec})
is quantized by $\frac{\pi}{2}$ in our regular
cubic lattice, the eigenvalue equation
possesses the symmetry $J \rightarrow 4+J$ and $J \rightarrow -J$.
This is because that
$J$ is a half-integer and
$Q({\bf 0}, -J)$ is equal to the transposed
matrix of $Q({\bf 0}, J)$.
Thus we only have to consider the cases :
$J=0, \ \frac{1}{2}, \ 1, \ \frac{3}{2}, \ 2$.
For each $J$, we find solutions of
(\ref{ee}) as in Table \ref{eigenvalues} . \\
\ \\
\begin{table}[t]
\label{eigenvalues}
\begin{center}
\renewcommand{\arraystretch}{1.2}
\begin{tabular}{|l||llllll|} \hline
{\it Spin Factor} & \multicolumn{6}{c|}{\it Eigenvalues} \\ \hline\hline
J=0 & {\bf 5} & 1 & 1 & 1 & $-1$ & $-1$ \\ \hline
J=1/2 & ${\bf 1+2\sqrt{2}}$ & ${\bf 1+2\sqrt{2}}$ & $1-\sqrt{2}$ &
        $1-\sqrt{2}$ & $1-\sqrt{2}$ & $1-\sqrt{2}$ \\ \hline
J=1 & {\bf 3} & {\bf 3} & {\bf 3} & $-1$ & $-1$ & $-1$ \\ \hline
J=3/2 & ${\bf 1+\sqrt{2}}$ & ${\bf 1+\sqrt{2}}$ & ${\bf 1+\sqrt{2}}$ &
        ${\bf 1+\sqrt{2}}$ & $1-2\sqrt{2}$ & $1-2\sqrt{2}$ \\ \hline
J=2 & 3 & 3 & 1 & 1 & 1 & $-3$ \\ \hline
\end{tabular}
\end{center}
\caption[Table]{Table of eigenvalues of the matrix $Q({\bf 0}, J)$}
\end{table}

One can see several degeneracies.
At $n$-th degenerate critical point,
the amplitude is expressed as a linear combination of
$n$ independent eigenvectors and
can survive as an excitation with long range correlation.
Only $(J, \kappa_c ^{-1})=(0,5), (\frac{1}{2}, 1+2 \sqrt{2}),
(1,3), (\frac{3}{2}, 1+\sqrt{2})$, (2, -3) and (2, 3)
have to be concerned for the convergence of the sum (\ref{sum}).
One can construct
rotationally invariant field theories at critical points
$(J, \kappa^{-1})=(0, 5)$, $(\frac{1}{2}, 1+ 2 \sqrt{2})$,
$(1, 3)$ and $(2, -3)$, if one tunes the diffusion constant
$\kappa$ to these critical points $\kappa_c$.
Those with single, double and triple
degeneracies give us Klein-Gordon, Dirac and
massive Chern-Simons fields, respectively.
We do not find any other critical points that
give continuum field theory. \\

First, we show the behavior of the free energy at
critical points which give field theories.
At $(J, \kappa^{-1})=(0, 5)$ and $(2, -3)$,
the critical behaviors belong to the same
universality class. The free energy around $(0,5)$ is
given
\begin{equation}
f(\kappa, 0) \simeq \int \frac{d^3 p}{(2 \pi)^3}
\log[{\bf p}^2+4s + O(p^4)], 
\label{k}
\end{equation}
where we introduce small parameter $s$
$$
s =  \frac{\kappa_c -\kappa}{\kappa_c} > 0 .
$$
This critical behavior belongs to the same universality class
of the Brownian motion. The correlation length diverges at
the critical point as $\sim s^{-\frac{1}{2}}$, thus
the critical exponent $ \nu = \frac{1}{2}$.
This critical point gives the Klein-Gordon theory as the
continuum limit.
The free energy near these two critical points behave as \\
$$
f(\kappa) \sim s^{\frac{3}{2}} \log s.
$$
This indicates that the order of the phase transition
is second and the critical
exponent $\alpha= \frac{1}{2}$. \\

At $(J, \kappa^{-1})=(\frac{1}{2}, 1+ 2 \sqrt{2})$ ,
the free energy is
\begin{equation}
f(\kappa,{\textstyle \frac{1}{2}})
\simeq - \int \frac{d^3 p}{(2 \pi)^3}
\log[{\bf p}^2 + s^2 + O(p^4)] \\
\sim s^{3} \log s.
\label{d1}
\end{equation}
The correlation length becomes $s^{-1}$, and then
$\nu=1$ and $\alpha=-1$.
The two component Dirac field is obtained
as the continuum limit at this critical point :
\begin{equation}
f(\kappa, {\textstyle \frac{1}{2}})
\simeq - \int \frac{d^3 p}{(2 \pi)^3}
{\rm Tr} \log[- i {\bf p} \cdot {\bf \sigma} -s],
\label{d2}
\end{equation}
where ${\bf \sigma}$ is the Pauli matrix
used as the gamma matrix in
three dimensions.
Other four components
correspond to heavy particles which does not affect
critical phenomena.

Here, we comment on a problem with fermion doubling on
a lattice. For $(J, \kappa^{-1})=(\frac{1}{2},
1+2 \sqrt{2})$, the matrix (\ref{mx})
has the only one massless excitation at $ {\bf p}={\bf 0}$.
The constant $\kappa^{-1}$ should be tuned
at $-1-2 \sqrt{2}$ but not at $1+2 \sqrt{2}$
to obtain massless
excitation at ${\bf p}=(\pi, \pi, \pi)$. The corresponding doubler
is too heavy at $\kappa^{-1}=1+2 \sqrt{2}$ to propagate,
as in the case of the Wilson fermion. Therefore,
no fermion doubling exists in our random walk model. \\

At $(J, {\kappa}^{-1}) = $  (1, 3),
the non-analytic part of the free energy at
$\kappa= 1+ \frac{s}{3}$ is
evaluated as follows:
\begin{equation}
f(\kappa, -1) \simeq \int \frac{d^3 p}{(2 \pi)^3}
 \log[\frac{16}{27} s {\bf p}^2 +
 \frac{64}{27} s^3 + O(p^4)]  \\
\sim s^{3} \log s.
\label{c1}
\end{equation}
The free energy can be written also in terms of
the following massive vector field
\begin{equation}
f(\kappa, -1) \simeq \int \frac{d^3 p}{(2 \pi)^3}
{\rm Tr} \log[i {\bf p} \cdot {\bf L} + 4 s],
\label{c2}
\end{equation}
where the matrix {\bf L} is a spin matrix with the
magnitude 1.  One can take
$(L_\mu)_{\nu \lambda} = i \epsilon_{\mu \nu \lambda}$.
Two critical exponents
$\alpha=-1$ and $\nu=1$ are same as those of the Dirac field.
The continuum limit is taken practically by
$s \rightarrow 0$ after the replacement
${\bf p} \rightarrow s^{\nu} {\bf p} $.
The model has infinite ultraviolet cutoff by this procedure. \\

Our method can construct the
propagator
$[1- \kappa Q(J, {\bf p})]^{-1}$, as well as the free energy.
Then, the Lagrangians of three types of rotationally invariant
field theories constructed above
are given for the Klein-Gordon field
\begin{equation}
{\cal L}_0 = \partial _\mu \phi ^{\ast} \partial _\mu \phi
+ m \phi^{\ast} \phi,
\end{equation}
the Dirac field
\begin{equation}
{\cal L}_{\frac{1}{2}} = \bar{\psi} (-i \gamma _\mu \partial _\mu
+ m) \psi,
\label{ld}
\end{equation}
and the massive U(1) Chern-Simons field
\begin{equation}
{\cal L}_1= -i \epsilon ^{\mu \nu \lambda}
A_\mu ^{\ast} \partial_\nu A_\lambda
 + m A_\mu ^{\ast} A_\mu.
\label{lv}
\end{equation}
Above these three types of critical theories agree with
the result given by the continuum path integral \cite{P}-\cite{T}. \\

The order of phase transition
at $(J, \kappa^{-1})=
(\frac{3}{2}, 1+\sqrt{2})$, and $(2, 3)$
seems higher than the second and then
the continuum limit might exist at those critical points.
One might expect spin $\frac{3}{2}$
field theory at $(\frac{3}{2}, 1+\sqrt{2})$.
Nevertheless, these continuum theories at
$J=\frac{3}{2}$ and $2$
possess no rotational symmetry. For example, the
free energy at $(J, \kappa^{-1})=(\frac{3}{2},
(1+\sqrt{2})(1+s))$ becomes
\begin{equation}
f(\kappa, {\textstyle \frac{3}{2}})
\simeq - \int \frac{d^3 p}{(2 \pi)^3}
\log[18(3-2 \sqrt{2}) \{ s^2 {\bf p}^2 + \frac{s^4}{2} + \frac{1}{4}
( p_2 ^2 p_3 ^2 + p_3 ^2 p_1 ^2 + p_1 ^2 p_2 ^2) + O(p^6) \}].
\label{rs}
\end{equation}
This shows no rotational invariance,
even though one takes the limit $s \rightarrow 0$.
Thus, the result for a spin higher than 1
does not agree with that obtained in the
continuum approach.
We can conclude that the continuum random walk method
derive wrong results for higher spin fields.

\section{Summary and Discussions}

We studied a random walk model
with a spin factor on a three dimensional lattice.
Relativistic field theories of Klein-Gordon,
Dirac and massive Chern-Simons fields have been
constructed by taking continuum limit
near critical points. It is found that there is no fermion
doubling in this model.
On the other hand, however, field
theory with a spin higher than 1
have not been obtained as a rotationally invariant model.

To check the universality of our results,
we have to work on some other lattice structures.
Here, we summarize the results we have obtained so far
in some other lattices such as triclinic and body
centered cubic lattices \cite{AIK}. All obtained results in a
regular cubic lattice are derived
in a body centered cubic lattice as well.
In any lattice we have examined, the Klein-Gordon and the
Dirac fields are obtained always. However, the massive
Chern-Simons field cannot be obtained in triclinic lattices. The triple
degeneracy of the matrix $Q({\bf 0}, 1)$
is split up in such lattice. This fact implies that
the construction of
spin 0 and spin 1/2 fields seems universal,
while that of spin 1 field is not universal.

\ \\

\section*{Acknowledgement}
The authors thank T. Hara, N. Sakai and G. Semenoff
for critical comments and helpful discussions.  \\
They are also grateful to J. Ambj\o rn, P. Kurzepa and
P. Orland for notifying them of references.

\end{document}